\documentclass[twocolumn]{article}

\usepackage[dvips]{graphicx}

\begin{document}

\begin{titlepage}

\hfill{LIGO TD-000012-R}

\begin{center}

\vfill
{\Large\bf Doppler-Induced Dynamics of Fields in Fabry-Perot 
Cavities with Suspended Mirrors}\footnote{published in 
{\it Applied Optics}, Vol. 40, No. 12, 20 April 2001, pp. 1942-1949}

\vspace{1cm}
{Malik~Rakhmanov}
\vspace{0.5cm}

{\it Physics Department, University of Florida, Gainesville, FL 32611}

\end{center}

\vfill
\begin{abstract}
The Doppler effect in Fabry-Perot cavities with suspended mirrors is
analyzed. Intrinsically small, the Doppler shift accumulates in the
cavity and becomes comparable to or greater than the line-width of the
cavity if its finesse is high or its length is large. As a result,
damped oscillations of the cavity field occur when one of the mirrors
passes a resonance position. A formula for this transient is derived.
It is shown that the frequency of the oscillations is equal to the
accumulated Doppler shift and the relaxation time of the oscillations
is equal to the storage time of the cavity. Comparison of the predicted 
and the measured Doppler shift is discussed, and application of the 
analytical solution for measurement of the mirror velocity is described.
\end{abstract}


\vfill
\end{titlepage}

\section*{Introduction}

Fabry-Perot cavities with the length of several kilometers are
utilized in laser gravitational wave detectors such as LIGO
\cite{Abramovici:1992}. The mirrors in these Fabry-Perot cavities 
are suspended from wires and therefore are free to move along the
direction of beam propagation. Ambient seismic motion excites the
mirrors, causing them to swing like pendulums with frequencies of 
about one hertz and amplitudes of several microns. To maintain the
cavity on resonance the Pound-Drever locking technique
\cite{Drever:1983} is used. During lock acquisition the mirrors
frequently pass through resonances of the cavity. As one of the mirrors
approaches a resonant position the light in the cavity builds up.
Immediately after the mirror passes a resonance position, a field
transient in the form of damped oscillations occurs. This transient
depends mostly on the cavity length, its finesse, and the relative
velocity of the mirrors. Thus, careful examination of the transient
reveals useful information about the cavity properties and the mirror
motion.

The oscillatory transient was observed in the past in several
experiments with high-finesse Fabry-Perot cavities. The oscillations
were recorded in the intensity of reflected light by Robertson {\it
et al.} \cite{Robertson:1989}. In this experiment the oscillations
were used for measurements of the storage time of a Fabry-Perot cavity
and its finesse. The oscillations were also observed in the intensity
of transmitted light by An {\it et al.} \cite{An:1995}. In this
experiment the oscillations were used for measurements of the cavity
finesse and the mirror velocity. The transient was also studied by
Camp {\it et al.} \cite{Camp:1995} for applications to cavity lock
acquisition. This time the oscillations were observed in the
Pound-Drever locking signal. Recently the transient has been revisited
by Lawrence {\it et al.} \cite{Lawrence:1999}. In this study both the
cavity length scans and the frequency scans were analyzed using all
three signals: the intensities of the reflected and transmitted fields
as well as the Pound-Drever signal.
 
Although the transient has been frequently observed in experiments,
its theory is far from being complete. It is known that the
oscillations in the transient appear through the beatings  of
different field components in the cavity. However, different authors
propose slightly different beat mechanisms \cite{An:1995,
Lawrence:1999}. Moreover, it is not understood why the  rate of the
oscillations always  increases in time, and what causes this
chirp-like behavior.

In this paper we show that the transient can be explained by the
Doppler effect which appears in a Fabry-Perot cavity with moving
mirrors. Intrinsically small, the Doppler shift is amplified by the
cavity and results in the modulation of the intracavity field. Based
on this observation, we derive a simple formula for the transient and
explain its chirp-like behavior. In this approach the frequency of
the oscillations can be easily  found from the cavity parameters and
the mirror velocity. The predictions based on the formula and
numerical simulations are compared with the measurements taken with
the 40m Fabry-Perot cavity of the Caltech prototype interferometer.
In both cases good agreements are found.

Currently the transient is studied in connection with locking of the
kilometer-sized Fabry-Perot cavities of LIGO interferometers
\cite{Yamamoto:1999}. The analysis presented in this  paper serves as
a basis for calculations of the cavity parameters in these studies.

\section{Reflection of Light off a Moving Mirror}

\begin{figure}[ht]
   \centering\includegraphics[width=0.4\textwidth]{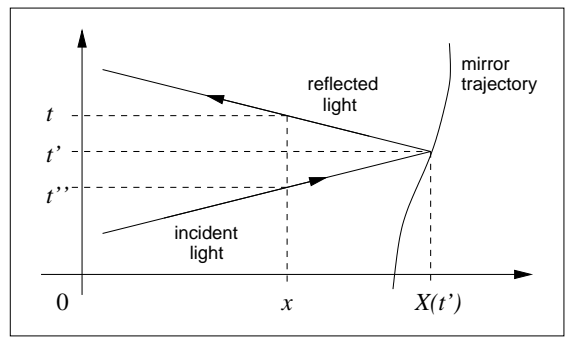}
   \caption{Reflection of light off a moving mirror.}
   \label{fig1}
\end{figure}

To set grounds for the analysis in this paper, consider a simple
process of reflection of light (electromagnetic wave) off a moving
mirror. Let the mirror be moving along the $x$-axis with an arbitrary
trajectory $X(t)$. Assume that the light is propagating along the
positive $x$-direction and is  normally incident on the mirror. The
wave-front of the  reflected wave observed at the location $x$ and
time $t$ is reflected  by the mirror at some earlier time $t'$ which,
according to Fig.~\ref{fig1}, satisfies the equation:
\begin{equation}\label{tp}
   c (t - t') = X(t') - x .
\end{equation}
This equation defines the time $t'$ as an implicit function of $x$
and $t$. 

Let the electric field of the incident and reflected waves be 
$\mathcal{E}_{\mathrm{in}}(x,t)$ and
${\mathcal{E}}_{\mathrm{ref}}(x,t)$. Due to continuity of the waves 
at the mirror surface, the two fields are related according to 
\begin{equation}\label{refWave}
   {\mathcal{E}}_{\mathrm{ref}}(x,t) = 
      {\mathcal{E}}_{\mathrm{in}}(X(t'), t') .
\end{equation}
For simplicity we assumed that the mirror is perfect (100\%
reflective), and no phase change occurs in the reflection.

Equations (\ref{tp}) and (\ref{refWave}) allow us to calculate the 
wave reflected by a mirror moving along an arbitrary trajectory.
Let the incident wave be plane and monochromatic,
\begin{equation}
   {\mathcal{E}}_{\mathrm{in}}(x,t) = 
      \exp \left\{ i(\omega t - k x) \right\} ,
\end{equation}
where $\omega$ is the frequency and $k$ is the wavenumber
($k=\omega/c$). Then the reflected wave is given by
\begin{equation}
   {\mathcal{E}}_{\mathrm{ref}}(x,t) = 
      \exp \left\{ i[\omega t' - k X(t')] \right\} .
\end{equation}
Substituting for $t'$ from Eq.~(\ref{tp}) we obtain that the 
electric field of the reflected wave is given by
\begin{equation}
   {\mathcal{E}}_{\mathrm{ref}}(x,t) = \exp \left\{ 
      i(\omega t + k x) \right\} \exp \left[-2 i k X(t')\right] .
\end{equation}

The extra phase, $-2kX(t')$, appears due to the continuity of the
waves at the mirror surface, and leads to two closely related
effects. On one hand, it gives rise to the {\it phase} shift of the
reflected wave  which appears because the mirror position is changing.
On the other hand, it gives rise to the {\it frequency} shift of the
reflected wave which appears because the mirror is moving. Indeed,
the frequency of the reflected wave can be found as
\begin{equation}\label{freqRWdef}
   \omega'(t) = \omega - 2 k \frac{d X}{d t'}
      \frac{\partial t'}{\partial t}.
\end{equation}
Note that $dX/dt$ is the instantaneous mirror velocity $v(t)$, and 
\begin{equation}\label{dtpdt}
   \frac{\partial t'}{\partial t} = \frac{c}{c + v(t')} ,
\end{equation}
which can be derived from Eq.~(\ref{tp}). Combining 
Eqs.~(\ref{freqRWdef}) and (\ref{dtpdt}), we obtain the 
formula for the frequency of the reflected wave:
\begin{equation}\label{freqRWexact}
   \omega'(t) = \frac{c - v(t')}{c + v(t')} \; \omega .
\end{equation}

At any given location the electric field oscillates at a very high 
frequency ($\mathcal{E} \propto e^{i\omega t}$). It is convenient 
to remove the high-frequency oscillating factor $e^{i\omega t}$ and
consider only the slowly varying part of the wave:
\begin{equation}
   E(t) \equiv {\mathcal{E}}(x,t) \; e^{-i\omega t} .
\end{equation}
The two amplitudes, $E_1(t)$ and $E_2(t)$, which correspond to the
same wave but defined at different locations, $x$ and $x'$, are
related:
\begin{equation}
   E_2(t) = E_1(t - L/c) \; e^{ -i k L},
\end{equation}
where $L$ is the distance between the two locations ($L=x'-x$).

We now obtain a formula for the reflection off the moving mirror in 
terms of the ``slowly-varying'' field amplitudes. This can be done  
by tracing the incident beam from the mirror surface back to the point
with the coordinate $x$:
\begin{equation}\label{refWaveX}
   {\mathcal{E}}_{\mathrm{ref}}(x,t) = 
      {\mathcal{E}}_{\mathrm{in}}(x, t'') ,
\end{equation}
where the time $t''$ is further in the past and according to 
Fig.~\ref{fig1} is given by 
\begin{equation}\label{tpp}
   t'' = 2 t' - t .
\end{equation}
Equations (\ref{refWaveX}) and (\ref{tpp}) lead to the following
relation between the field amplitudes:
\begin{equation}
   E_{\mathrm{ref}}(t) = E_{\mathrm{in}}(t'') \; 
      \exp\{- 2ik [X(t') - x] \} .
\end{equation}
This formula is used below for calculations of fields in Fabry-Perot
cavities with moving mirrors.

For non-relativistic mirror motion ($|v|\ll c$) the frequency of the 
reflected light can be approximated as
\begin{equation}\label{freqRWapprox}
    \omega'(t) \approx \left[1 - 2 \frac{v(t')}{c} \right] \omega ,
\end{equation}
which differs from the exact formula, Eq.~(\ref{freqRWexact}),
only in the second order in $v/c$.

\section{Doppler Shift in Fabry-Perot Cavities}
\subsection{Critical Velocity}

Fabry-Perot cavities of laser gravitational-wave detectors are very
long and have mirrors that can move. The Doppler shift in such
cavities can be described as follows. Let the cavity length be
$L$ and the light transit time be $T$:
\begin{equation}
   T = \frac{L}{c}.
\end{equation}
Assume that one of the mirrors is moving with the constant velocity $v$. 
Then the frequency of light reflected off the moving mirror is Doppler 
shifted, and the shift in one reflection is 
\begin{equation}\label{dw}
   \delta \omega \equiv \omega' - \omega = - 2 kv .
\end{equation}
Subsequent reflections make this frequency shift add, forming the
progression:
\begin{equation}\label{prog}
   \delta \omega, \; 2 \delta \omega, \; 3 \delta \omega, \ldots .
\end{equation}
Therefore, the Doppler shift of light in the cavity accumulates with
time. 

A suspended mirror in such cavities moves very little. Its largest
velocity is typically of the order of a few microns per second. 
The corresponding Doppler shift is of the order of a few hertz, which
is very small compared to the laser frequency $2.82\times10^{14}$ 
Hz for an infra-red laser with wavelength $\lambda = 1.06 \mu$m. 
However, the line-width of the long Fabry-Perot cavities of the laser 
gravitational wave detectors is also very small, typically of the 
order of 100 Hz. Therefore, the small Doppler shift, as it accumulates 
with time, can easily exceed the line-width.

The characteristic time for light to remain in the cavity is the
storage time, which is defined as $1/e$-amplitude folding time:
\begin{equation}\label{storTime}
   \tau = \frac{2T}{|\ln (r_a r_b)|} ,
\end{equation}
where $r_a$ and $r_b$ are the amplitude reflectivities of the cavity 
mirrors. Then the Doppler shift accumulated within the storage time is
\begin{equation}
   |\delta \omega| \frac{\tau}{2T} = \omega \frac{v\tau}{cT} .
\end{equation}
It becomes comparable to the line-width of the cavity if the relative 
velocity of the mirrors is comparable to the critical velocity defined 
as
\begin{equation}
   v_{\mathrm{cr}} = \frac{\lambda}{2 \tau {\mathcal{F}}} 
      \approx \frac{\pi c \lambda}{4 L {\mathcal{F}}^2} ,
\end{equation}
where ${\mathcal{F}}$ is the finesse of the cavity:
\begin{equation}
   {\mathcal{F}} = \frac{\pi \sqrt{r_a r_b}}{1 - r_a r_b} .
\end{equation}
Note that the mirror moving with the critical velocity passes the 
width of a resonance within the storage time. These qualitative
arguments show that the Doppler effect becomes significant if 
the time for a mirror to move across the width of a resonance is 
comparable to or less than the storage time of the cavity.

\subsection{Equation for Fields in a Fabry-Perot Cavity}

The response of Fabry-Perot cavities is usually expressed in terms of
amplitudes of the electro-magnetic field circulating in the cavity.
The equation for the dynamics of this field can be derived as follows.
Assume, for simplicity, that one of the mirrors (input mirror) is at 
rest and the other (end mirror) is freely swinging. Let the trajectory 
of this mirror be $X(t)$. It is convenient to separate the constant
and the variable parts of the mirror trajectory:
\begin{equation}
    X(t) = L + x(t) .
\end{equation}
In Fabry-Perot cavities of gravitational wave detectors $L$ is of the 
order of a few kilometers and $x$ is of the order of a few microns.
Without loss of generality we can assume that the cavity length $L$ is 
equal to an integer number of wavelengths and therefore $e^{-2ikL}=1$.

\begin{figure}[ht]
   \centering\includegraphics[width=0.45\textwidth]{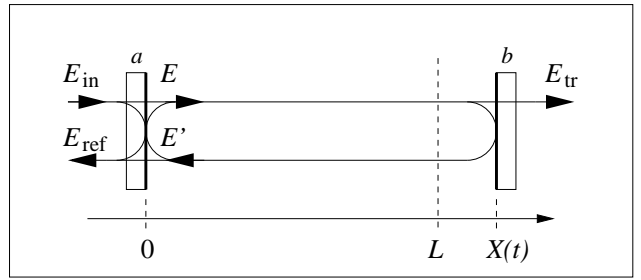}
   \caption{Schematic diagram of a Fabry-Perot cavity with
   a moving mirror.}
   \label{fig2}
\end{figure}

Let the amplitude of the input laser field be $E_{\mathrm{in}}(t)$ and
the amplitudes of the fields inside the cavity be $E(t)$ and $E'(t)$, 
both defined at the reflective surface of the input mirror as shown in
Fig.~\ref{fig2}. Then the equation for reflection off the end
mirror can be written as follows
\begin{equation}\label{reflEnd}
   E'(t) = - r_b E(t - 2T) \exp \left[ -2ik x(t - T) \right] ,
\end{equation}
where $r_b$ is the amplitude reflectivity of the end mirror. A similar
equation can be written for the reflection off the front mirror:
\begin{equation}\label{reflFront}
   E(t) = - r_a E'(t) + t_a E_{\mathrm{in}}(t) ,
\end{equation}
where $t_a$ is the transmissivity of the front mirror.

Finally, the amplitudes of the transmitted and the reflected field are
given by
\begin{eqnarray}
   E_{\mathrm{tr}}(t)  & = & t_b E(t - T) , \label{Etr} \\
   E_{\mathrm{ref}}(t) & = & r_a E_{\mathrm{in}}(t) + t_a E'(t) ,
      \label{Eref}
\end{eqnarray}
where $t_b$ is the transmissivity of the end mirror. Note that the 
reflected field is a superposition of the intracavity field leaking 
through the front mirror and the input laser field reflected by the 
front mirror, as shown in Fig.~\ref{fig2}.

It is convenient to reduce Eqs.~(\ref{reflEnd}) and
(\ref{reflFront}) to one equation with one field:
\begin{equation}\label{eqE}
   E(t) = t_a E_{\mathrm{in}}(t) + r_a r_b E(t - 2T) 
      \exp \left[-2ik x(t - T)\right] .
\end{equation}
Further analysis of field dynamics in the Fabry-Perot cavities is
based on this equation.

\section{Transient due to Mirror Motion}

The mirrors in Fabry-Perot cavities of laser gravitational wave
detectors are suspended from wires and can swing like pendulums 
with frequencies of about 1 Hz. The amplitude of such motion is of 
the order of a few microns. During the swinging motion, the mirrors
frequently pass through resonances of the cavity. Each passage
through a resonance gives rise to the field transient in the form of
damped oscillations.  Such a transient can be described in terms of
the complex amplitude of the cavity field as follows. For the entire
time when the mirror moves through the width of a resonance (a few
milliseconds), its velocity  can be considered constant, and its
trajectory can be approximated as linear: $x=vt$. Often the
amplitude of the incident field is constant: $E_{\mathrm{in}}(t)=A$. 
Then amplitude of the intracavity  field satisfies the equation:
\begin{equation}\label{AmpE}
   E(t) = t_a A + r_a r_b E(t - 2T) \exp \left[ 
      -2ik v \cdot (t - T) \right] ,
\end{equation}
which is a special case of equation (\ref{eqE}).

\begin{figure}[ht]
   \centering\includegraphics[width=0.45\textwidth]{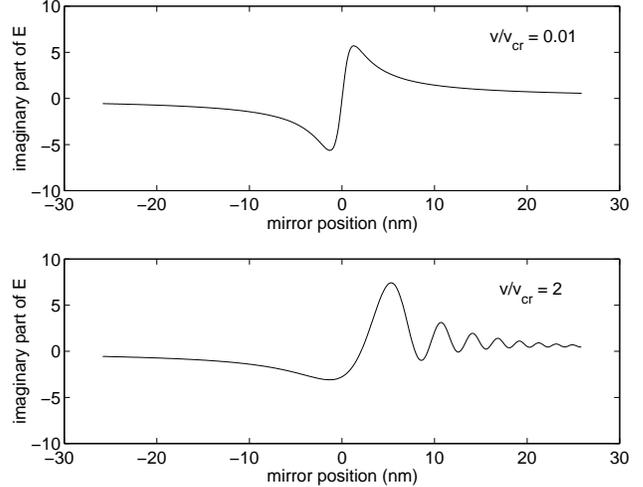}
   \caption{Modeled response of the LIGO 4km Fabry-Perot cavity 
   (finesse 205). The two curves correspond to the slow and the fast 
   motion of the mirror ($v_{\mathrm{cr}} = 1.48 \times 10^{-6}$ m/s).} 
   \label{fig3}
\end{figure}

Numerical solutions of this equation can be easily obtained on the
computer. Examples of the numerical solution with the parameters of 
LIGO 4km Fabry-Perot cavities are shown in Fig.~\ref{fig3}.

Such numerical solutions provide an accurate description for the field 
transient but give little insight into the physics of the process. 
Therefore, it is worthwhile to obtain an approximate analytical
solution for this equation.

\subsection{Approximate Solution for the Transient}

An approximate solution can be derived as follows. A general solution 
of Eq.~(\ref{AmpE}) can be represented as a sum:
\begin{equation}\label{CpD}
   E(t) = C(t) + D(t) .
\end{equation}
Here $C(t)$ is a particular solution of the non-homogeneous equation 
and $D(t)$ is a general solution of the homogeneous equation:
\begin{eqnarray}
   C(t) - r_a r_b C(t - 2T) \exp \left[-2ik v \cdot 
       (t - T)\right] & = & t_a A ,\label{EqC} \\
   D(t) - r_a r_b D(t - 2T) \exp \left[-2ik v \cdot 
       (t - T)\right] & = & 0 .\label{EqD}
\end{eqnarray}
Both amplitudes, $C(t)$ and $D(t)$, change very little during one 
round-trip. In the case of $C$-field the approximation
$C(t-2T) \approx C(t)$ yields the solution:
\begin{equation}\label{Ct}
   C(t) \approx \frac{t_a A}{1 - r_a r_b \exp (- 2ik vt)} ,
\end{equation}
which is generally known as the adiabatic field. (Here we also made 
the approximation: $v \cdot (t - T) \approx vt$.) The adiabatic
component was introduced empirically by Yamamoto \cite{Yamamoto:1999}.

In the case of $D$-field the approximation $D(t - 2T) \approx D(t)$
yields only a trivial solution: $D(t)=0$. Fortunately, the equation
for $D$-field can be solved exactly. A trial solution for $t > 0$ is
\begin{equation}
   D(t) = D_0 (r_a r_b)^{t/2T} \exp [i \phi(t)] ,
\end{equation}
where $D_0$ is the value of $D$-field at time $t = 0$ and $\phi(t)$ is
an arbitrary phase. Then Eq.~(\ref{EqD}) reduces to the equation 
for the phase: 
\begin{equation}
   \phi(t) = \phi(t - 2T) - 2kv \cdot (t - T) .
\end{equation}
Its solution, up to an additive constant, is
\begin{equation}
   \phi(t) = - \frac{k v}{2T} t^2 .
\end{equation}
Thus, we obtain the solution for $D$-field:
\begin{equation}\label{Dt}
   D(t) = D_0 \exp \left( - \frac{t}{\tau} -
      i \frac{kv}{2T} t^2 \right) ,
\end{equation}
where $\tau$ is the cavity storage time defined in
Eq.~(\ref{storTime}). This expression is valid for $t > 0$ and describes 
the phase modulation of the cavity field due to the Doppler effect.
The constant $D_0$ can be found from the asymptotic behavior of the
field \cite{Rakhmanov:PhD} and is given by
\begin{equation}\label{D0}
   D_0(kv) = t_a A \; \left( \frac{i\pi}{2kvT} \right)^{\frac{1}{2}} 
      \exp \left( \frac{iT}{2 kv \tau^2} \right) .
\end{equation}
Equation (\ref{Dt}) shows that $D$-field is oscillating with the 
frequency which linearly increases with time:
\begin{equation}\label{Omega}
   \Omega(t) \equiv \left| \frac{d\phi}{dt} \right| = 
      \frac{k |v|}{T} t .
\end{equation}
Note that the frequency of the oscillations is equal to the
accumulated Doppler shift:
\begin{equation}
   \Omega(t) = |\delta \omega| \frac{t}{2T} ,
\end{equation}
where $\delta \omega$ is the frequency shift which occurs in one
reflection off the moving mirror, Eq.~(\ref{dw}).

Combining the above results we obtain the approximate formula for 
the transient:
\begin{eqnarray}
   E(t) & \approx & \frac{t_a A}{1 - r_a r_b \exp (- 2ik vt)} 
          \nonumber \\
        & & + D_0(kv) \exp \left( - \frac{t}{\tau} - 
            i \frac{kv}{2T} t^2 \right) \label{Ecomp} .
\end{eqnarray}
Thus the transient, which occurs during a passage of the mirror
through a resonance, is caused by the Doppler effect amplified by the
cavity. The frequency of oscillations linearly increases in time with
the rate proportional to the mirror velocity.

Comparison of the approximate analytical solution given by 
Eq.~(\ref{Ecomp}) with the numerical simulations based on 
Eq.~(\ref{AmpE}) shows that the two solutions agree very well in the
region past the resonance ($t \gg T$). However, the two solutions
differ substantially in the region near the center of the resonance
($t \approx 0$). This is because the center of the resonance is the
boundary of the validity of the approximate analytical solution.

The above analysis leads to the following explanation of the
oscillatory transient. As the mirror approaches the resonance position
($x = 0$), the light is rapidly building in the cavity. At the time
when the mirror passes the center of the resonance, a substantial
amount of light accumulates in the cavity. From this moment on, the
light stored in the cavity ($D$-component) decays according to the
exponential law, and its frequency is  continuously shifting due to
the Doppler effect. At the same time there is a constant influx of
the new light from the laser ($C$-component). The new light is not
affected by the Doppler shift and therefore evolves according to the
usual adiabatic law.

\subsection{Observation of the Transient via Beats}

The small frequency shifts of the light circulating in the cavity are
usually observed through beats. There are several beat mechanisms
which take place in Fabry-Perot cavities with moving mirrors. Here we
describe the three most frequently occurred beat mechanisms in detail.

The Doppler-induced oscillations of the intracavity field can be
observed in the intensity of the transmitted field. The above
analysis shows that the Doppler effect gives rise to {\it phase}
modulation of $D$-field. As a result, the cavity field $E$, which
is the sum of $D$ and $C$ fields, becomes {\it amplitude} modulated.
This amplitude modulation can be observed as the intensity modulation
of the field transmitted through the cavity. According to 
Eqs.~(\ref{Etr}) and (\ref{CpD}) the intensity of the transmitted
field is proportional to
\begin{eqnarray}\label{ampMod}
   |E(t)|^2 
      & \approx & |C(t)|^2 + |D(t)|^2 \nonumber \\
      & & + 2 \; {\mathrm{Re}} \{ C(t)^* \; D(t) \} ,
\end{eqnarray}
where an asterisk stands for complex conjugation. 
Note that neither $|C(t)|^2$ nor $|D(t)|^2$ are oscillating functions.
Therefore, the oscillations come from the last term, which represents
a beating between $D$ and $C$-components of the intracavity field.

Similarly, the oscillations of the intracavity field can be observed
in the intensity of the reflected field. According to 
Eqs.~(\ref{reflEnd}) and (\ref{Eref}) the amplitude of the reflected
field can be found as
\begin{equation}
   E_{\mathrm{ref}}(t) = [ (r_a^2 + t_a^2) 
      E_{\mathrm{in}}(t) - t_a E(t) ]/r_a .
\end{equation}
For high-finesse cavities ($r_a \approx 1$) with low losses 
($r_a^2 + t_a^2 \approx 1$) the complex amplitude of the reflected
field can be approximated as
\begin{equation}
   E_{\mathrm{ref}}(t) \approx E_{\mathrm{in}}(t) - t_a E(t) .
\end{equation}
Then the intensity of the reflected light is given by
\begin{eqnarray}
   |E_{\mathrm{ref}}(t)|^2 & \approx & 
      |E_{\mathrm{in}}(t)|^2 + t_a^2 |E(t)|^2 \nonumber \\
      & & - 2 t_a {\mathrm{Re}} \{ E_{\mathrm{in}}(t)^* E(t)\}.
\end{eqnarray}
The second term in the right hand side of this equation  represents
the amplitude  modulation of the intracavity field as described in
Eq.~(\ref{ampMod}). The last term represents a beating of the
intracavity field transmitted through the front mirror and the input
laser field promptly reflected by the front mirror. Both terms give
rise to the oscillations in the intensity of the reflected field. 
Therefore the decay of the reflected intensity is described by the 
double exponential function with two decay times, $\tau$ and $\tau/2$,
as was noticed by Robertson {\it et al.} \cite{Robertson:1989}.

The oscillations can also be observed in the Pound-Drever signal,
which requires optical sidebands be imposed on the light incident  on
the cavity. In this case the signal is obtained from beating of the
carrier reflected field with the sideband reflected fields. Since the
carrier field propagates in the cavity, it becomes Doppler-shifted
due to the  motion of the cavity mirrors. The sideband fields are
promptly reflected by the front mirror of the cavity. Therefore their
amplitudes are  proportional to the amplitude of the incident carrier
field. Then the signal can be approximated by the formula:
\begin{equation}\label{PDdef}
   V(t) = - {\mathrm{Im}} \{ e^{i\gamma} 
      E_{\mathrm{in}}(t)^* E(t) \} ,
\end{equation}
where $\gamma$ is the phase of a local oscillator in the optical 
heterodyne detection.

If the amplitude of the input laser field is constant
($E_{\mathrm{in}}(t) = A$) then the Pound-Drever signal becomes a 
linear function of the cavity field:
\begin{eqnarray}
   V(t) & = & - A \; {\mathrm{Im}} \{ e^{i\gamma} E(t) \}
             \label{PDsig}\\
   & \approx & - A \; {\mathrm{Im}} \{ e^{i\gamma} [C(t) + D(t)] \} .
\end{eqnarray}
Since $C$-component is a monotonic function, the oscillations come
from $D$-component only. Unlike the signals derived from the
intensity of the transmitted  or reflected fields, the Pound-Drever
signal is linearly proportional to the amplitude of the intracavity
field and therefore presents a direct way to observe the oscillations.

\section{Experimental Analysis of the Transient}

\begin{figure}
   \centering\includegraphics[width=0.45\textwidth]{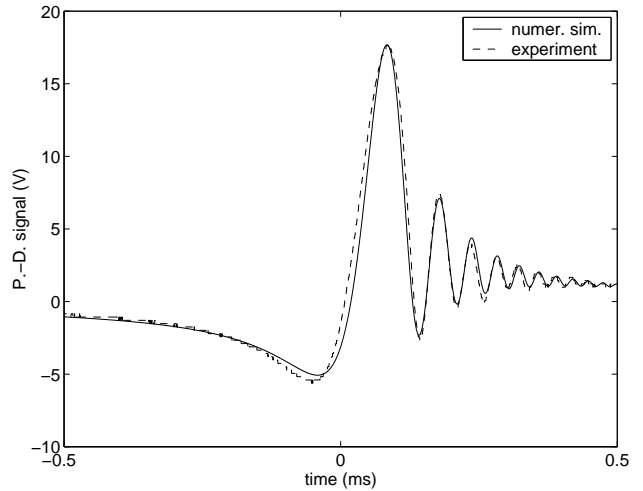}
   \caption{Transient response of the Fabry-Perot cavity of the
   Caltech 40m prototype interferometer ($v/v_{\mathrm{cr}}=1.93$).}
   \label{fig4}
\end{figure}

The measurements of the oscillatory transient analyzed in this paper
were taken with the 40m Fabry-Perot cavity of the LIGO-prototype
interferometer at Caltech. The experimental setup was previously
described by Camp {\it et al.} \cite{Camp:1995}. Figure \ref{fig4}
shows the Pound-Drever signal of the 40m Fabry-Perot cavity recorded
with a digital oscilloscope (dashed line). The theoretical prediction
shown in the same Figure (solid line) is obtained by numerical
simulations of the intracavity field using Eq.~(\ref{AmpE}).
After adjustment of the demodulation phase ($\gamma\approx-0.28$ rad)
a good agreement between the theoretical and the experimental curves
was achieved. It is important to note that the mirror velocity 
($v \approx 5.5 \times 10^{-6}$ m/s) used for the numerical
simulations was not a fit parameter. It was obtained from the 
interpolation of the mirror trajectory using the optical vernier
technique \cite{Rakhmanov:optVer}.

The formula for the transient, Eq.~(\ref{Ecomp}), can be used 
for extracting the cavity parameters from the Pound-Drever signal. 
In such an analysis it is convenient to remove the adiabatic 
component from the Pound-Drever signal. The result is the function 
very similar to $D(t)$, which is given by
\begin{eqnarray}
   V_D(t) & = & - A |D_0| \exp \left( -\frac{t - t_0}{\tau} 
                \right) \nonumber \\
          &   & \times \sin \left[ \gamma + \delta -
                \frac{kv}{2T} (t - t_0)^2 \right] \label{VD} .
\end{eqnarray}
Here we introduced $t_0$, the time when the mirror passes a center
of the resonance, and $\delta = \arg D_0$. The measured Pound-Drever
signal (with the adiabatic component removed) and the theoretical 
prediction based on this formula are shown in Fig.~\ref{fig5}
(upper plot).

\subsection{Measurement of the Cavity Finesse}

The oscillatory transient can be used for measurements of the cavity 
finesse. The present approach is based on the exponential decay of the
Pound-Drever signal. The finesse can be found by studying the absolute
value of the adjusted Pound-Drever signal:
\begin{equation}
   |V_D(t)| \propto \exp(- t/\tau) .
\end{equation}
Indeed, by fitting the exponential function to the envelope of the 
oscillations $|V_D(t)|$, one can find the storage time of the cavity, 
$\tau$, and therefore its finesse:
\begin{equation}
   {\mathcal{F}} = \frac{\pi}{2 \sinh (T/\tau)} .
\end{equation}
Applied to the data shown in Fig.~\ref{fig4}, this method yields the 
following value for the finesse of the Caltech 40m Fabry-Perot cavity:
\begin{equation}
   {\mathcal{F}} = 1066 \pm 58 .
\end{equation}
This result is close to the one previously obtained from the 
measurement of the mirror reflectivities (${\mathcal{F}}\approx1050$).
The present approach to measure the cavity storage time is similar to
the one described by Robertson {\it et al} \cite{Robertson:1989}.

\subsection{Measurement of the Mirror Velocity}

The oscillatory transient can also be used for measurements of the
mirror velocity. The present approach is based on the linear shift 
of the frequency of the Pound-Drever signal. The velocity can be 
found by studying either the peaks or the zero crossing of the
adjusted Pound-Drever signal, $V_D(t)$. 

Let the times for the zero crossings be $t_n$, where $n$ is integer. 
The values for $t_n$ are defined by the functional form of the
adjusted Pound-Drever signal, Eq.~(\ref{VD}), and are given by
\begin{equation}
   \frac{k v}{2T} (t_n - t_0)^2 = \pi n + \gamma + \delta .
\end{equation}
This relation depends on the demodulation phase $\gamma$, which is 
not always known in the experiment. However, the difference:
\begin{equation}\label{identity}
   \frac{k v}{2T} \left[ (t_{n + 1} - t_0)^2 - 
      (t_n - t_0)^2 \right] = \pi ,
\end{equation}
does not depend on the demodulation phase and therefore is more 
suitable for this analysis. Define the spacings between the zero 
crossings, $\Delta t_n$, and the positions of their midpoints, 
$\bar{t}_n$, as follows:
\begin{eqnarray}
   \Delta t_n & = & t_{n + 1} - t_n , \\
   \bar{t}_n  & = & \frac{1}{2} \left( t_n + t_{n + 1} \right) .
\end{eqnarray}
Then the ``average'' frequency of the oscillations of $V_D(t)$, can
be defined as 
\begin{equation}
   \bar{\nu}_n = \frac{1}{2 \Delta t_n} .
\end{equation}
Using the identity, Eq.~(\ref{identity}), we can show that the
average frequencies satisfy the equation:
\begin{equation}\label{linFreq}
   \bar{\nu}_n = \frac{v}{\lambda T} (\bar{t}_n - t_0) .
\end{equation}
This equation is a discrete analog of the continuous evolution,
Eq.~(\ref{Omega}). 

If the times $t_n$ correspond to the peaks and not the zero crossings 
of the signal, the predicted average frequency becomes
\begin{equation}
   \bar{\nu}_n = \frac{v}{\lambda T} (\bar{t}_n - t_0) + 
      \delta \bar{\nu}_n ,
\end{equation}
where $\delta \bar{\nu}_n$ is a small correction which accounts for
the exponential decay of the signal present in Eq.~(\ref{VD}). The 
correction can be found from Eq.~(\ref{VD}) using a perturbation 
expansion in powers of $T/\tau$. In the lowest order, it is given by
\begin{equation}
   \delta \bar{\nu}_n = - \frac{4 \lambda v T (\bar{t}_n - t_0)^2}
      {\pi^2 \tau \left[ 16 v^2 (\bar{t}_n - t_0)^4 - 
      \lambda^2 T^2 \right]} .
\end{equation}
Such a correction becomes significant only if the oscillations are
close to being critically damped.

\begin{figure}
   \centering\includegraphics[width=0.45\textwidth]{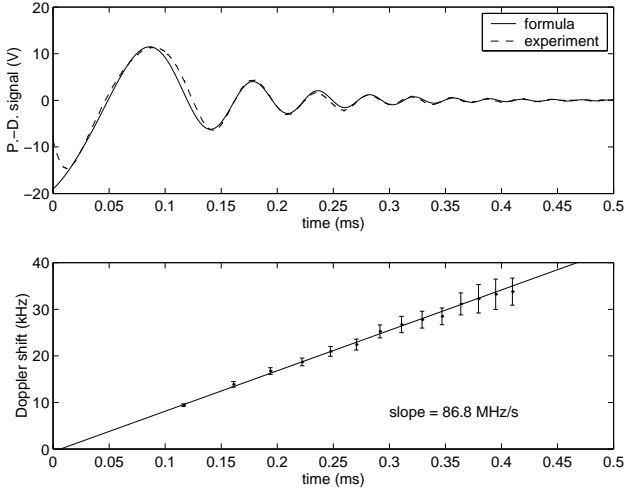}
   \caption{Upper diagram: Theoretical prediction (solid line)
   and the measurement (dashed line) of the adjusted Pound-Drever 
   signal. Lower diagram: measured Doppler shift $\bar{\nu}_n$ and 
   the linear fit $\bar{\nu}(t)$.}
   \label{fig5}
\end{figure}

In general the zero crossings can be affected by the subtraction of
the adiabatic component. Therefore, we prefer to use the peaks of the
signal. The peak-positions $t_n$ are found from the measured
Pound-Drever signal, which is shown in Fig.~\ref{fig5} (upper plot). 
Since the oscillations are far from being critically damped,the 
correction $\delta \bar{\nu}_n$ can be neglected. In this experiment,
the first order correction is much less than the error in
determination of the average frequencies. As a result the measured
values of the average frequencies $\nu_n$ appear very close to the
linear function, Eq.~(\ref{linFreq}). This can also be seen in
Fig.~\ref{fig5} (lower plot). Therefore, we can apply a linear fit
to the data:
\begin{equation}
   \bar{\nu}(t) = a t + b ,
\end{equation}
where $a$ and $b$ are the slope and the intercept of the linear function.
The least square adjustment of the fit gives the following values for
these parameters:
\begin{eqnarray}
   a & = & (86.8 \pm 0.6) \times 10^{6}\ {\mathrm{Hz/s}} ,\\
   b & = & (-0.5 \pm 1.0) \times 10^{3}\ {\mathrm{Hz}} .
\end{eqnarray}
The slope is related to the mirror velocity, and the intercept is 
related to the time when mirror passes through the center of the
resonance:
\begin{eqnarray}
   v   & = & \lambda T a ,\\
   t_0 & = & - b/a.
\end{eqnarray}
From these relations we obtain
\begin{eqnarray}
   v   & = & (5.7 \pm 0.4) \times 10^{-6}\ {\mathrm{m/s}} ,\\
   t_0 & = & (0.6 \pm 1.2) \times 10^{-5}\ {\mathrm{s}} .
\end{eqnarray}
The errors are due to uncertainty in the peak positions, which are
limited in this measurement by the resolution of the oscilloscope.

\section*{Conclusion}

The Doppler effect in Fabry-Perot cavities with suspended mirrors can
be significant and manifests itself in the oscillations of the field
transient, which can be directly observed via the Pound-Drever signal.
The transient can be used for accurate measurements of the cavity
finesse and the mirror velocities. Implemented in real-time computer
simulations the formula for the transient can be used in lock
acquisition algorithms.

The analysis presented in this paper explains the chirp-like
behavior of the transient and leads to a simple formula for its
frequency. However, the approximate analytical solution given in this
paper describes only the ringdown part of the transient. The buildup
part is yet to be explained. Also it is not clear at the present time
why oscillations always appear after the mirror passes the center of
the resonance and not before.

\section*{Acknowledgment}

I thank Guoan Hu for assistance with the experiment, and Hiro Yamamoto
and Matt Evans for helpful discussions. I also thank Barry Barish and
other scientists of LIGO project: Rick Savage, David Shoemaker and
Stan Whitcomb for their suggestions about the manuscript during the
draft process. Finally, I thank David Reitze and David Tanner of the
University of Florida for the discussions of the transient and their
comments on the paper. This research was supported by the National
Science Foundation under Cooperative Agreement PHY-9210038.

\end{document}